\documentclass{aastex6}

\graphicspath{{./}{figures/}}

\shorttitle{Large-scale cosmic-ray anisotropies above 4 E\lowercase{e}V}
\shortauthors{The Pierre Auger Collaboration}

\begin{document}

\title{Large-scale cosmic-ray anisotropies above 4 E\MakeLowercase{e}V measured by\\ the Pierre Auger Observatory}

\slugcomment{Published in Astrophys. J as DOI: 10.3847/1538-4357/aae689}


\author{
A.~Aab\altaffilmark{75},
P.~Abreu\altaffilmark{67},
M.~Aglietta\altaffilmark{50,49},
I.F.M.~Albuquerque\altaffilmark{19},
J.M.~Albury\altaffilmark{12},
I.~Allekotte\altaffilmark{1},
A.~Almela\altaffilmark{8,11},
J.~Alvarez Castillo\altaffilmark{63},
J.~Alvarez-Mu\~niz\altaffilmark{74},
G.A.~Anastasi\altaffilmark{42,43}
% ...
}


\altaffiltext{1}{Centro At\'omico Bariloche and Instituto Balseiro (CNEA-UNCuyo-CONICET), San Carlos de Bariloche, Argentina}
% ...
\altaffiltext{8}{Instituto de Tecnolog\'\i{}as en Detecci\'on y Astropart\'\i{}culas (CNEA, CONICET, UNSAM), Buenos Aires, Argentina}
% ...
\altaffiltext{11}{Universidad Tecnol\'ogica Nacional -- Facultad Regional Buenos Aires, Buenos Aires, Argentina}
\altaffiltext{12}{University of Adelaide, Adelaide, S.A., Australia}
% ...
\altaffiltext{19}{Universidade de S\~ao Paulo, Instituto de F\'\i{}sica, S\~ao Paulo, SP, Brazil}
% ...
\altaffiltext{42}{Gran Sasso Science Institute, L'Aquila, Italy}
\altaffiltext{43}{INFN Laboratori Nazionali del Gran Sasso, Assergi (L'Aquila), Italy}
% ...
\altaffiltext{49}{INFN, Sezione di Torino, Torino, Italy}
\altaffiltext{50}{Osservatorio Astrofisico di Torino (INAF), Torino, Italy}
% ...
\altaffiltext{63}{Universidad Nacional Aut\'onoma de M\'exico, M\'exico, D.F., M\'exico}
% ...
\altaffiltext{67}{Laborat\'orio de Instrumenta\c{c}\~ao e F\'\i{}sica Experimental de Part\'\i{}culas -- LIP and Instituto Superior T\'ecnico -- IST, Universidade de Lisboa -- UL, Lisboa, Portugal}
% ...
\altaffiltext{74}{Instituto Galego de F\'\i{}sica de Altas Enerx\'\i{}as (I.G.F.A.E.), Universidad de Santiago de Compostela, Santiago de Compostela, Spain}
\altaffiltext{75}{IMAPP, Radboud University Nijmegen, Nijmegen, The Netherlands}
% ...

\collaboration{The Pierre Auger Collaboration}


\vfill

\newpage

\begin{abstract}

We present a detailed study of the large-scale anisotropies of cosmic rays with energies above 4~EeV measured using the Pierre Auger Observatory. For the energy bins [4,\,8]~EeV and $E\geq 8$~EeV, the most significant signal is a dipolar modulation in right ascension at energies above 8~EeV, as previously reported.  In this paper we further scrutinize the highest-energy bin by splitting it into three energy ranges. 
 We find that the amplitude of the dipole increases with energy above 4~EeV. The growth can be fitted with a  power law with index $\beta=0.79\pm 0.19$. The directions of the dipoles are consistent with an extragalactic origin of these anisotropies at all the energies considered. Additionally we have estimated the quadrupolar components of the anisotropy: they are not statistically significant. We discuss the results in the context of the predictions from different models for the distribution of ultrahigh-energy sources and cosmic magnetic fields.
\end{abstract}


\section{Introduction} \label{sec:intro}

The distribution of the arrival directions of cosmic rays (CR) with ultrahigh energies is expected to play a major role in the quest to unveil the origin of these particles. Hints of anisotropies at intermediate ($\sim 10^\circ$--~$20^\circ$) angular scales have been reported at the highest energies, above $\sim 40$~EeV (where 1~${\rm EeV}\equiv 10^{18}$~eV), by  searching  for a localized excess in the cosmic-ray flux or for    correlations with catalogs of candidate  populations of astrophysical sources \citep{apj2015,sb2017,tahotspot}. None of these results has a large-enough statistical significance to claim a detection.
At $E\ge 8$~EeV,  a first-harmonic modulation in right ascension was detected  with a significance of more than 5.2$\sigma$ \citep{science}. The amplitude of the three-dimensional dipolar component that was determined in this energy bin is  $\sim 6.5$\%, with its direction lying $\sim 125^\circ$ away from the Galactic center direction and hence indicating an extragalactic origin for this flux. 

 The observation of a significant dipole, together with the lack of significant anisotropies at small angular scales, implies that the Galactic and/or extragalactic magnetic fields have a non-negligible effect on ultrahigh-energy cosmic-ray (UHECR) trajectories. This is in fact expected in scenarios with  mixed composition where the CRs  are heavier for increasing energies, in agreement with the trends in the composition that have been inferred for energies  above a few EeV \citep{augerxmax,sein,hybrid,delta}. Extragalactic magnetic fields can significantly spread the arrival directions of heavy CR nuclei up to the highest energies observed, even for nearby extragalactic sources, washing out small-scale anisotropies while still leading to anisotropies at large (and eventually intermediate) angular scales.\footnote{The root-mean square deflection of a particle of charge $Z$ and energy $E$ in a homogeneous turbulent magnetic field with root mean square amplitude $B$ and coherence length $l_c$ is $\delta_{rms}\simeq 30^\circ (B/{\rm nG})(4Z \,{\rm EeV}/E)\sqrt{l_c/{\rm Mpc}}\sqrt{L/10\,{\rm Mpc}}$. For instance,  oxygen nuclei with 30~EeV coming from a distance $L\simeq 10$~Mpc are deflected by about $30^\circ$ for an extragalactic field of 1~nG, which is consistent with the bounds from cosmic background radiation  and Faraday rotation measures \citep{durrer}.} The Galactic magnetic field is also expected to further modify the arrival directions of extragalactic CRs, affecting both the amplitude and the direction of the dipolar contribution to their flux and also inducing some higher multipolar components when the deflections become sizable. It is not yet clear whether the dipolar anisotropy observed arises from the diffusive propagation from powerful sources in a few nearby galaxies or is instead reflecting the known anisotropy in the distribution of galaxies within few hundred Mpc \citep{gi80,be90,hmr14,hmr15}. A detailed study of the amplitude and phase of the dipole as a function of energy, as well as the possible emergence of structures at  smaller angular scales, should shed light on the distribution of the sources and on the strength and structure of the magnetic fields responsible for the deflections.

We present here an extension of the analysis of the large-scale anisotropies measured by the Pierre Auger Observatory for energies above 4~EeV. We obtain both the dipolar and quadrupolar components in the two energy ranges  that were discussed  in \citet{lsa2015,science}, i.e.  [4,\,8]~EeV and $E\ge 8$~EeV. We further analyze the bin above 8~EeV by splitting it into three so as to explore how the amplitude and phase of the dipole changes with energy. We then discuss the results obtained in the frame of scenarios proposed for the origin of the large-scale anisotropies.

\section{The Observatory and the dataset}
The Pierre Auger Observatory, located near Malarg\"ue, Argentina  \citep{augerobs}, has an array of surface detectors (SD)  that covers an area of 3000~km$^2$. The array contains 1660 water-Cherenkov detectors, 1600 of which are deployed on a triangular grid with 1500~m spacing with the remainder on a lattice of 750~m covering 23.5~km$^2$. The array is  overlooked by 27 fluorescence telescopes (FD) that are used to monitor the longitudinal development of the air showers during moonless and clear nights, with a duty cycle of about $13\%$. The SD has a duty cycle of about $100\%$ so that it provides the vast majority of the events, and it is hence adopted for the present study. The energy of these events is calibrated using hybrid events measured simultaneously by both SD and FD.

The dataset analyzed in this work is the same one as that considered in \citet{science}, including events from the SD array with 1500~m separation detected from 2004 January 1 up to 2016 August 31.
We retain events with  zenith angles up to $80^\circ$ and energies in excess of 4~EeV, for which the array is fully efficient over  the full zenith angle range considered.\footnote{The  smaller but denser sub-array with 750~m spacing among detectors is fully efficient down to $\sim 0.3$~EeV for events with $\theta<55^\circ$. The large-scale anisotropy results that can be obtained using it will be presented elsewhere.} The events with zenith angles
$\theta \le 60^\circ$, referred to as  vertical, have a different reconstruction and calibration from the ones having $60^\circ < \theta \le 80^\circ$, referred to as inclined events. The energies of the  vertical sample are corrected for atmospheric effects \citep{jinst17}, since, otherwise, systematic modulations of the rates as a function of the hour of the day or of the season, and hence also as a function of right ascension, could be induced.
These effects arise from the dependence on the atmospheric conditions of the longitudinal and lateral attenuation of the electromagnetic component of the extended air showers. The energies are also corrected for geomagnetic effects \citep{geomagnetic} since,
otherwise, systematic modulations in the azimuthal distribution could result. Results from the inclined sample, for which the signal from the muonic component of the
extended air showers is dominant, have negligible dependence on the atmospheric effects while the geomagnetic field effects are already accounted for in the reconstruction.
We include events for which at least 5 of the 6 neighboring stations to the one with the largest signal are active at the time at which the event is recorded \citep{science}. Adopting  this cut, the  total integrated exposure of the array in the period considered is 76,800\,km$^2$\,sr\,yr.
Selecting events with zenith angles up to $80^\circ$ allows us to explore all the directions with declinations between $-90^\circ \le \delta \le 45^\circ$, covering $85\%$ of the sky. The total number of recorded events above the energy  threshold of 4~EeV is 113,888.

\section{Large-scale anisotropy results}
Above full trigger efficiency for the SD array, which is achieved for $E\ge 4$~EeV when zenith angles up to  $80^\circ$ are considered, the systematic effects relevant for the distributions of the events in right ascension ($\alpha$) and in the azimuth angle ($\phi$) are well under control (see Section~\ref{systematics}). One can hence obtain a reliable estimate of the three-dimensional dipole components, and eventually also higher multipoles, from the Fourier analysis in these two angular coordinates after including appropriate weights to account for known systematic effects  \citep{lsa2015}. The method adopted, based on the harmonic analyses on $\alpha$ and $\phi$, does not require to have a detailed knowledge of the distribution of the event directions that would be expected for an isotropic flux after all detector, calibration and atmospheric  effects are included.
It thus has the advantage of being largely insensitive to possible distortions in the zenith-angle distribution of the events, such as those that could result from a difference in the relative energy calibration of the vertical and inclined samples.

The harmonic amplitudes of order $k$ are given by 
\begin{equation}
    a_k^x = \frac{2}{\mathcal{N}}\sum_{i=1}^{N}w_i\cos(k x_i)\ \ \ ,\ \ \ 
	b_k^x = \frac{2}{\mathcal{N}}\sum_{i=1}^{N}w_i\sin(k x_i),
    \label{harmonic}
\end{equation}
with $x=\alpha$ or $\phi$.
The sums run over the number of events $N$ in the energy range  considered and the normalization factor is $\mathcal{N} = \sum_{i=1}^N w_i$. The weight factors $w_i$ take into account the modulation in the exposure due to dead times of the detectors and also account for the effects of the tilt of the array, which on average is inclined $0.2^\circ$ towards $\phi_0\simeq -30^\circ$ (being the azimuth measured anti-clockwise from the East direction). The weights, which are of order unity, are given by \citep{lsa2015}
\begin{eqnarray}
w_i &=& \left[\Delta N_{\rm cell}(\alpha_i^0)(1+0.003\tan\theta_i\cos(\phi_i-\phi_0))\right]^{-1},
\end{eqnarray}
with the factor $\Delta N_{\rm cell}(\alpha_i^0)$ being the relative variation of the total number of active detector cells for a given right ascension of the zenith of the observatory $\alpha^0$,  evaluated at the time $t_i$ at which the $i$-th event is detected,  $\alpha_i^0=\alpha^0(t_i)$, and $\phi_i$ and $\theta_i$ are the azimuth and the zenith angle of the event, respectively.

The amplitude $r_k^x$ and phase $\varphi_k^x$ of the event rate modulation are given by
\begin{equation}
r_k^x = \sqrt{(a_k^x)^2+(b_k^x)^2},\hspace{20pt}\varphi_k^x=\frac{1}{k}\arctan\frac{b_k^x}{a_k^x}  .
\end{equation}\label{eq6}
The probability that an amplitude equal to or larger than $r_k^x$ arises as a fluctuation from an isotropic distribution is given by $P(\geq r_k^x)=\exp(-\mathcal{N}(r_k^x)^2/4)$ \citep{linsley}. 

In this work we will focus on the first two harmonics. Note that the first-harmonic amplitudes, corresponding to $k=1$, are the only ones present when the flux is purely dipolar. The second order harmonics, with $k=2$, are also relevant in the case of a flux with a non-vanishing quadrupolar contribution.

\subsection{Harmonic analysis in right ascension and azimuth}
Table~\ref{tab:radist} contains the results of the first and second harmonic analyses in right ascension for the two energy bins that were considered in previous publications, [4,\,8]~EeV and $E\ge 8$~EeV. The statistical uncertainties in the harmonic coefficients are $\sqrt{2/\mathcal{N}}$.  No significant harmonic amplitude is observed in the first bin, while for energies above 8~EeV the $p$-value for the first harmonic is $2.6 \times 10^{-8}$. The results for the first harmonics were already presented in \cite{science}.

\begin{table}[ht]
\centering
\caption{Results of the first and second harmonic analyses in right ascension.}
\begin{tabular}{c c c c c c c c}
\hline\hline
    Energy [EeV] &events & $k$ & $a_k^\alpha$ &  $b_k^\alpha$ & $r_k^\alpha$ & $\varphi_k^\alpha [^\circ]$ & $P(\ge r_k^\alpha)$ \\
\hline
 4 - 8  & 81,701 &1 &  $0.001 \pm 0.005$ & $0.005 \pm 0.005$ & 0.005 & $80\pm 60$ & 0.60 \\
 & & 2 &  $-0.001 \pm 0.005$ & $0.001 \pm 0.005$ & 0.002  & $70\pm 80$ & 0.94 \\
 \hline
$\geq$ 8  &32,187 & 1 & $-0.008 \pm 0.008$ & $0.046 \pm 0.008$ & 0.047 & $100\pm 10$ & $2.6 \times 10^{-8}$  \\
& & 2 & $0.013 \pm 0.008$ & $0.012 \pm 0.008$ & 0.018 & $21\pm 12$ & 0.065 \\
\end{tabular}
\label{tab:radist}
\end{table}

\begin{figure}[h]
\centering
\includegraphics[scale=1.03]{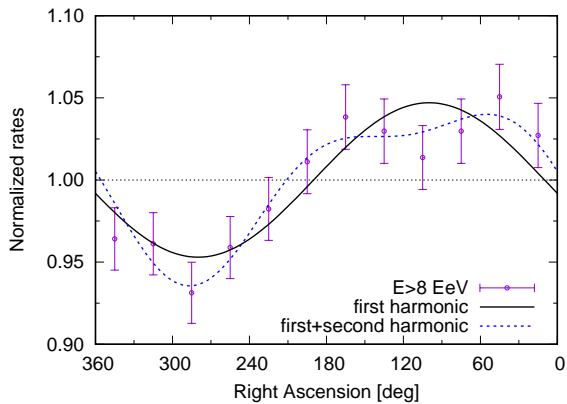}
\caption{Distribution in right ascension of the normalized rates of events with energy above 8~EeV. The black (solid)  and the blue (dashed) lines show the distributions obtained from the weighted Fourier analysis corresponding to a first harmonic ($\chi^2/{\rm dof}=1.02$, for 10 degrees of freedom) and first plus second harmonics ($\chi^2/{\rm dof}=0.44$, for 8 degrees of freedom), respectively. }
\label{fig:distra}
\end{figure}

In Fig.~\ref{fig:distra}, we display the distribution in right ascension of the normalized rates in the  energy bin $E\ge 8$~EeV.  We also show with a black solid line the first-harmonic modulation obtained through the Rayleigh analysis and the distribution corresponding to  a first plus second harmonic, with the amplitudes and phases reported in Table
\ref{tab:radist}. 

In Table~\ref{tab:phidist}, we report the results of the harmonic analysis in the azimuth angle. The $a_1^\phi$ amplitudes, that give a measure of the difference between the flux coming from the East  and that coming from the West, integrated over time, should vanish if there are no spurious modulations affecting the azimuth distribution. The values obtained  are in fact compatible with zero in the two bins.
The $b_1^\phi$ amplitudes, that give a measure of the flux modulation in the North-South direction, can be used to estimate the component of the CR dipole along the Earth rotation axis. The most significant amplitude is obtained for energies between 4 and 8~EeV and is $ b_1^\phi =-0.013 \pm 0.005$, corresponding to an excess CR flux from the South, that has a chance probability to arise from an isotropic distribution of $0.009$. Regarding the second harmonic, none of the amplitudes found are significantly different from zero.

\begin{table}[ht]
\centering
\caption{Results of the first and second harmonic analyses in azimuth.}
\begin{tabular}{c c c c c c }
\hline\hline
   Energy [EeV] & $k$ & $a_k^\phi$ &  $b_k^\phi$ & $P(\ge \mid a_k^\phi \mid)$ & $P(\ge \mid b_k^\phi \mid)$ \\
\hline
 4 - 8  & 1 &  $-0.010 \pm 0.005$ & $-0.013 \pm 0.005$ & 0.045 & 0.009 \\
 & 2 &  $0.002 \pm 0.005$ & $-0.002 \pm 0.005$ & 0.69 & 0.69 \\
 \hline
$\geq$ 8  & 1 & $-0.007 \pm 0.008$ & $-0.014 \pm 0.008$ & 0.38 & 0.08  \\
 & 2 & $-0.002 \pm 0.008$ & $0.006 \pm 0.008$ & 0.80 & 0.45 \\
\end{tabular}
\label{tab:phidist}
\end{table}

\begin{figure}[h]
\centering
\includegraphics[scale=.53]{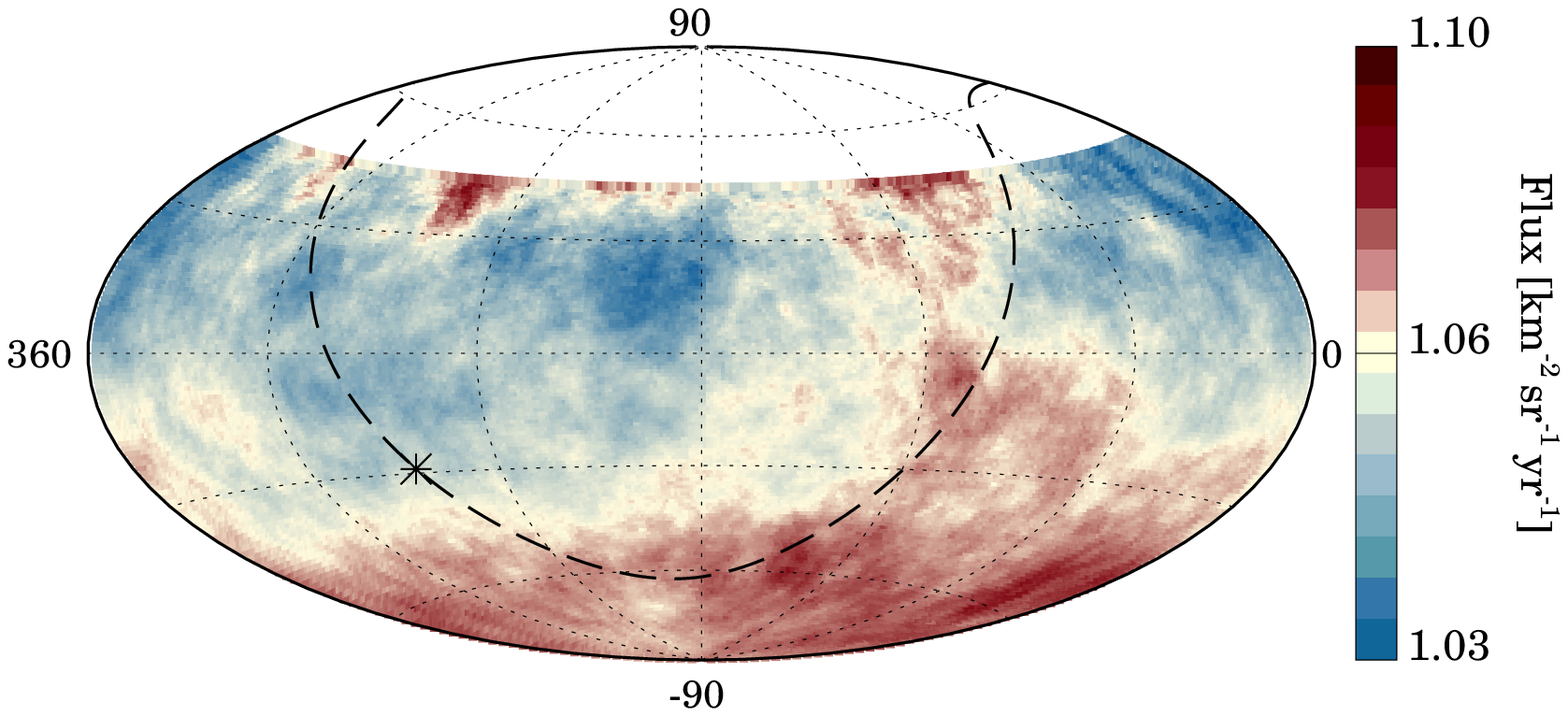}
\includegraphics[scale=.53]{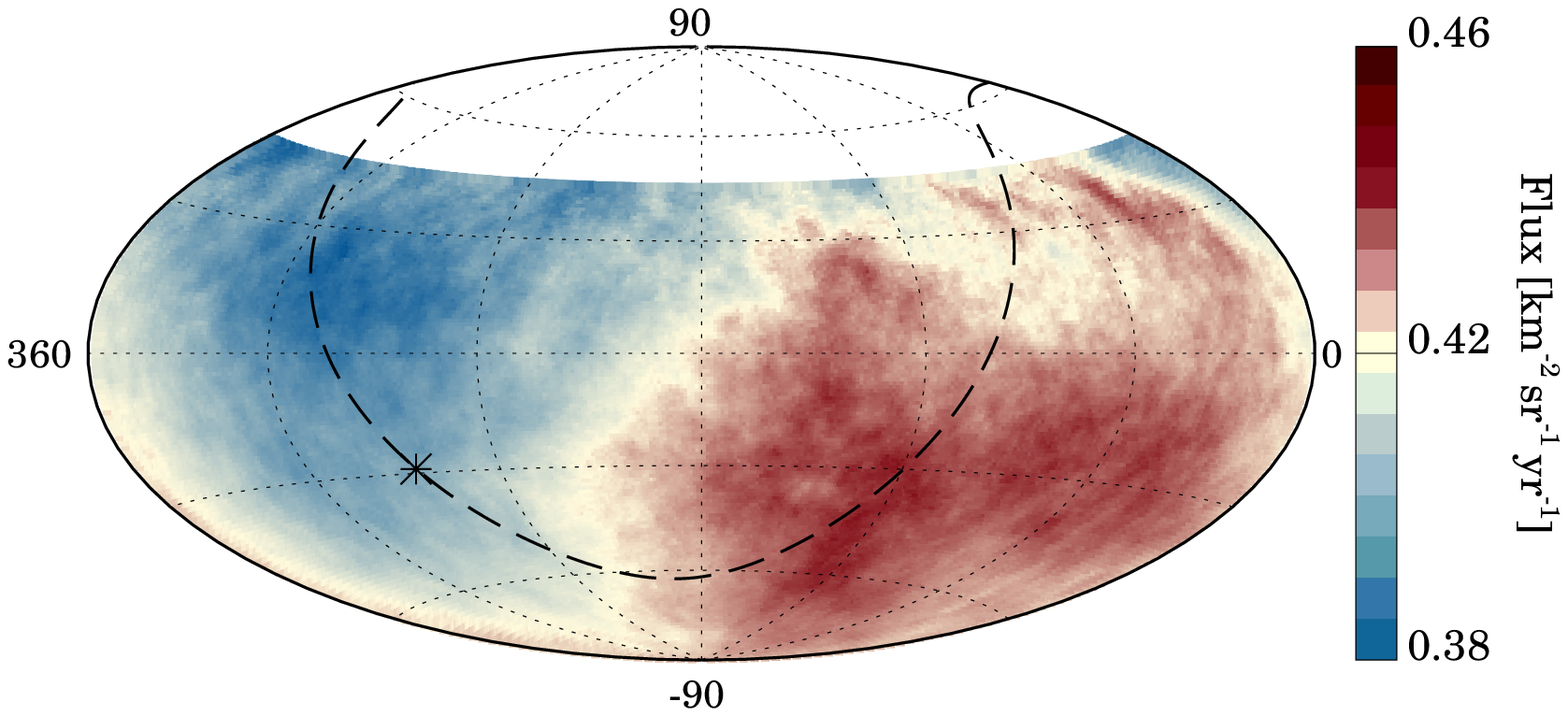}
\caption{Maps in equatorial coordinates of the CR flux, smoothed in windows of 45$^\circ$, for the energy bins [4,\,8]~EeV (left) and $E\ge 8$~EeV (right). The Galactic plane is represented with a dashed line and the Galactic center is indicated with a star.}
\label{fig:mapsa4}
\end{figure}

 Figure~\ref{fig:mapsa4} displays the maps, in equatorial coordinates, of the exposure-weighted average of the flux inside a  top-hat window of  radius $45^\circ$, so as to better
 appreciate the large-scale features, for the  energy bins  [4,\,8]~EeV and $E \ge 8$~EeV.
An excess in the flux from the southern directions is the predominant feature at energies between 4 and 8~EeV, while above 8~EeV the excess comes from a region with right ascensions close to $100^\circ$, with a corresponding deficit in the opposite direction, in accordance with the results from the harmonic analyses in right ascension and azimuth. 

\begin{table}[ht]
\centering
\caption{Results of the first-harmonic analysis in right ascension in the three bins above 8~EeV.}
\begin{tabular}{ c c c c c c c}
\hline\hline
Energy [EeV] & events &  $a_1^\alpha$ & $b_1^\alpha$ & $r_1^\alpha$ & $\varphi_1^\alpha$ [$^\circ$] &
$P(\ge r_1^\alpha)$ \\
\hline
8 - 16   &24,070& $-0.011 \pm 0.009$ & $\phantom{-}0.044 \pm 0.009$ & 0.046 & $104\pm 11$ & $3.7\times10^{-6}$ \\ 
16 - 32  &6,604& $\phantom{-}0.007 \pm 0.017$ & $\phantom{-}0.050 \pm 0.017$ & 0.051 & $82\pm 20$ & $0.014$ \\ 
$\ge 32$ &1,513 & $-0.03 \pm 0.04$ & $0.05 \pm 0.04$ & 0.06 & $115\pm 35$ & $0.26$\\ 

\end{tabular}
\label{tab:raharmonics_a8}
\end{table}
 
\begin{table}[h]
\centering
\caption{Results of the first-harmonic analysis in azimuth in the three bins above 8~EeV.}
\begin{tabular}{ c c c c c c }
\hline\hline
Energy [EeV]  & $a_1^\phi$ & $b_1^\phi$ & $P(\ge |a_1^\phi|)$ &  $P(\ge |b_1^\phi|)$ \\
\hline
8 - 16  & $-0.013 \pm 0.009$ & $-0.004 \pm 0.009$ & 0.15 & 0.66\\
16 - 32 & $\phantom{-}0.003 \pm 0.017$ & $-0.042 \pm 0.017$ & 0.86 & 0.013\\
$\ge 32$ & $\phantom{-}0.05 \pm 0.04$ & $-0.04 \pm 0.04$ & 0.21 & 0.32\\
\end{tabular}
\label{tab:phiharmonics_a8}
\end{table}

Given the significant first-harmonic modulation in right ascension that was found in the bin with $E\ge8$~EeV, we now divide this higher energy bin into three to study the possible energy dependence of this signal. For this,  we use energy boundaries  scaled by factors of two, i.e. considering the bins [8,\,16]~EeV, [16,\,32]~EeV and $E\ge32$~EeV. 
Table~\ref{tab:raharmonics_a8} reports the results for the right ascension analysis in these new energy bins. The $p$-values for the first-harmonic modulation in right ascension are $3.7\times 10^{-6}$ in the [8,\,16]~EeV range, 0.014 in the [16,\,32]~EeV bin and 0.26 for energies above 32~EeV. 
 Table~\ref{tab:phiharmonics_a8} reports the results for the corresponding azimuth analysis in these new energy bins. 

\subsection{Reconstruction of the CR dipole}

We now convert the harmonic coefficients in right ascension and in azimuth into anisotropy parameters on the sphere, assuming first that the dominant component of the anisotropy is the dipole  $\vec{d}$. The flux distribution can then be parametrized as a function of the CR arrival direction ${\hat u}$ as
\begin{equation}
\label{phidip}
\Phi({\hat u})=\Phi_0(1+{\vec d}\cdot {\hat u}).
\end{equation}
In this case, the amplitude of the dipole component along the rotation axis of the Earth, $d_z$, that in the equatorial plane, $d_\perp$, and the right ascension and declination of the dipole direction, ($\alpha_d,\delta_d)$, are related to the first-harmonic amplitudes in right ascension and azimuth through \citep{lsa2015}
\begin{eqnarray}
d_z &\simeq& \frac{b_1^\phi}{\cos\ell_{\rm obs}\langle\sin\theta\rangle},\nonumber\\
d_\perp &\simeq& \frac{r_1^\alpha}{\langle\cos\delta\rangle},\nonumber \\
\alpha_d &=& \varphi_1^\alpha ,\nonumber \\
\delta_d &=& \arctan\left(\frac{d_z}{d_\perp}\right),\label{eqdip}
\end{eqnarray}
where $\langle\cos\delta\rangle\simeq 0.7814$ is the mean cosine   of the  declinations of the events, $\langle\sin\theta\rangle\simeq 0.6525
$ the mean sine of the event zenith angles, and $\ell_{\rm obs}\simeq -35.2^\circ$ is the latitude of the Observatory.   Note that, as is well known, when the coverage of the sky is not complete a  coupling between the reconstructed multipoles can occur. The dipole parameters inferred from this set of relations can thus receive extra contributions from higher-order multipoles, something that will be explicitly checked in the next subsection in the case of a non-negligible quadrupolar contribution to the flux.

\begin{table}[ht!]
\centering
\caption{Three-dimensional dipole reconstruction for energies above 4~EeV. We show the results obtained for the two bins previously reported \citep{science}, i.e. between 4 and 8~EeV and above 8~EeV, as well as dividing the high-energy range into three bins.}
\begin{tabular}{c c c c c c c }
\hline\hline
    \multicolumn{2}{c}{Energy [EeV]} & $d_\perp$ &   $d_z$ & $d$ & $\alpha_d$ [$^\circ$] & $\delta_d$ [$^\circ$] \\
   interval & median & & & & & \\
\hline
 \rule{0pt}{3ex}  4 - 8 & 5.0 &  $0.006^{+0.007}_{-0.003}$  & $-0.024 \pm 0.009$ & $0.025^{+0.010}_{-0.007}$ & $80 \pm 60$ & $-75^{+17}_{-8}$ \\
 $\geq 8$  & 11.5 & $0.060^{+0.011}_{-0.010}$  & $-0.026 \pm 0.015$ & $0.065^{+0.013}_{-0.009}$ & $100 \pm 10$ & $-24^{+12}_{-13}$ \\
 \hline
 \rule{0pt}{3ex} 8 - 16  & 10.3 &  $0.058^{+0.013}_{-0.011}$ &  $-0.008 \pm 0.017$ & $0.059^{+0.015}_{-0.008}$ & $104 \pm 11$ & $-8^{+16}_{-16}$ \\
16 - 32  & 20.2 & $0.065^{+0.025}_{-0.018}$  & $-0.08 \pm 0.03$ & $0.10^{+0.03}_{-0.02}$ & $82 \pm 20$ & $-50^{+15}_{-14}$ \\
$\ge 32$  & 39.5 & $0.08^{+0.05}_{-0.03}$  & $-0.08 \pm 0.07$ & $0.11^{+0.07}_{-0.03}$ & $115 \pm 35$ & $-46^{+28}_{-26}$ 
\end{tabular}
\label{tab:3da8}
\end{table}

In the two upper rows of Table~\ref{tab:3da8}, we show the reconstructed dipole components for the energy bins previously studied,  [4,\,8]~EeV and $E\ge8$~EeV. The results for the three new bins above 8 EeV are reported in the lower three rows. The uncertainties in the amplitude and phase correspond to the $68\%$ confidence level of the marginalized probability distribution functions.

In Table~\ref{tab:3da8} a growth of the dipolar amplitude $d$ with increasing energies is observed. Adopting for the energy dependence of the dipole amplitude a power-law behavior 
$d(E) = d_{10} \times (E/10~\rm{EeV})^\beta$, we perform a maximum-likelihood fit to the values measured in the four bins above 4~EeV. We consider a likelihood function ${\cal L }( d_{10}, \beta) =\prod_{i=1}^4 f (\vec{d}_i;d_{10}, \beta)$, where in each energy bin $f$ is given by a three-dimensional Gaussian for the dipole vector $\vec d=d(E)(\cos\delta\cos\alpha,\cos\delta\sin\alpha,\sin\delta)$, centered at the measured dipole values and with the  dispersions $\sigma_x=\sigma_y=\sqrt{2/\mathcal{N}}/\langle\cos\delta\rangle$ and $\sigma_z=\sqrt{2/\mathcal{N}}/(\langle\sin\theta\rangle\cos\ell_{\rm obs})$, marginalized over the angular variables $\alpha$ and $\delta$.  
The fit leads to a reference amplitude $d_{10} = 0.055 \pm 0.008$ and a power-law index $\beta = 0.79 \pm 0.19$.\footnote{Regarding the goodness of the fit, we have checked that, for a model in which the dipole amplitude follows the power-law obtained, a better agreement than  the one found with the actual data is expected to result in about 50\% of the realizations.}
A fit with an energy-independent dipole amplitude ($\beta =0$) is disfavored at the level of 3.7$\sigma$ by a likelihood-ratio test.
\begin{figure}[ht]
\centering
\includegraphics[scale=.85]{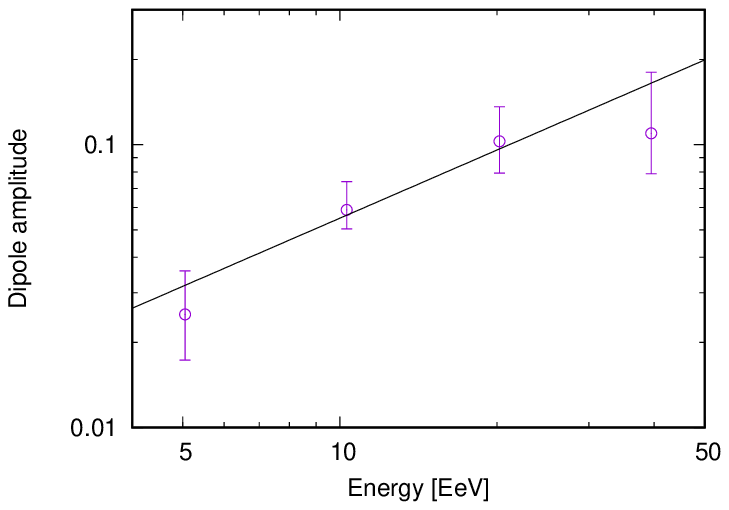}
\includegraphics[scale=.58]{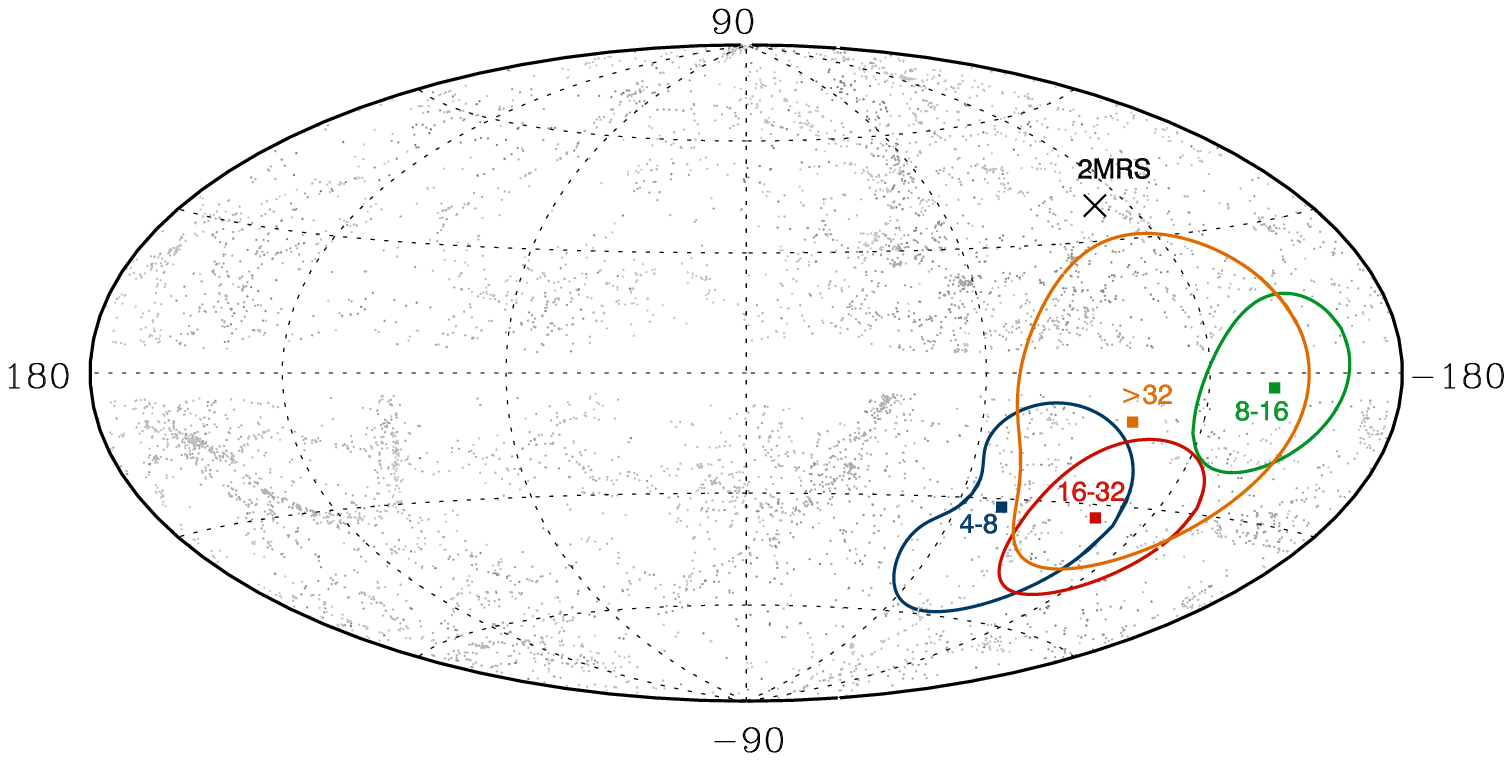}
\caption{Evolution with energy of the amplitude (left panel) and direction (right panel) of the three-dimensional dipole determined in different energy bins above 4~EeV. In the sky map in Galactic coordinates of the right panel the dots represent the direction towards the galaxies in the 2MRS catalog that lie within 100~Mpc and the cross indicates the direction towards the flux-weighted dipole inferred from that catalog.}
\label{fig:amp-fasea4}
\end{figure}

The left panel of Fig.~\ref{fig:amp-fasea4} shows the amplitude of the dipole as a function of the energy, with the data points centered at the median energy in each of the four  bins above 4~EeV, as well as the power-law fit. The right panel is a map, in Galactic coordinates,  showing the 68\% CL sky regions for the dipole direction in the same bins. They all point towards a similar region of the sky, and in order of increasing energies they are centered at Galactic coordinates ($\ell,b)=(287^\circ,-32^\circ)$, $(221^\circ,-3^\circ)$, $(257^\circ,-33^\circ)$ and $(259^\circ,-11^\circ)$, respectively. With the present accuracy no clear trend in the change of the dipole direction
as a function of energy  can be identified. 
In the background of Fig.~\ref{fig:amp-fasea4}, we indicate with dots the location of the observed galaxies from the 2MRS catalog that lie within 100~Mpc and 
also show with a cross the reconstructed 2MRS flux-weighted dipole direction \citep{erdogdu}, which could be expected to be related to the CR dipole direction if the galaxies were to trace the distribution of the UHECR sources and the effects of the magnetic field deflections were ignored.

\begin{figure}[h]
\centering
\includegraphics[scale=.45]{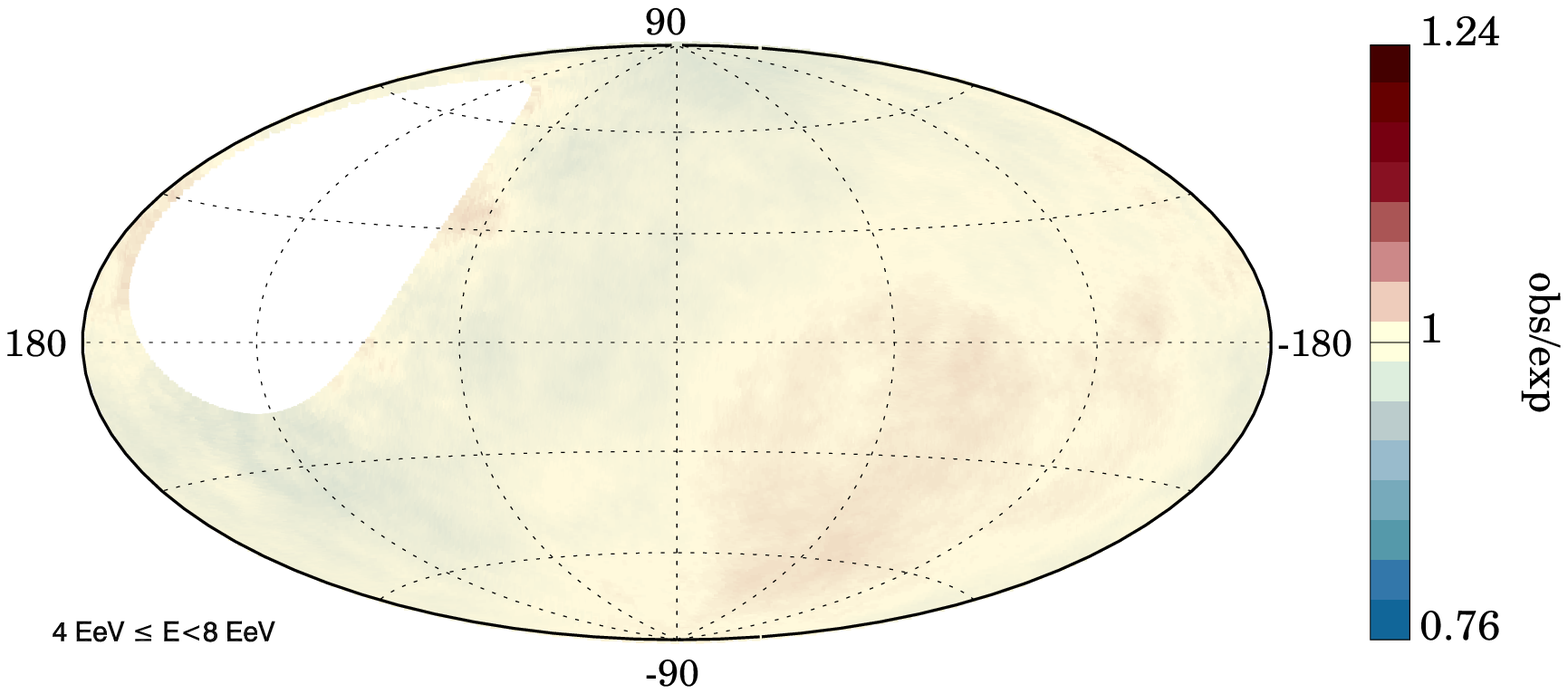}
\includegraphics[scale=.45]{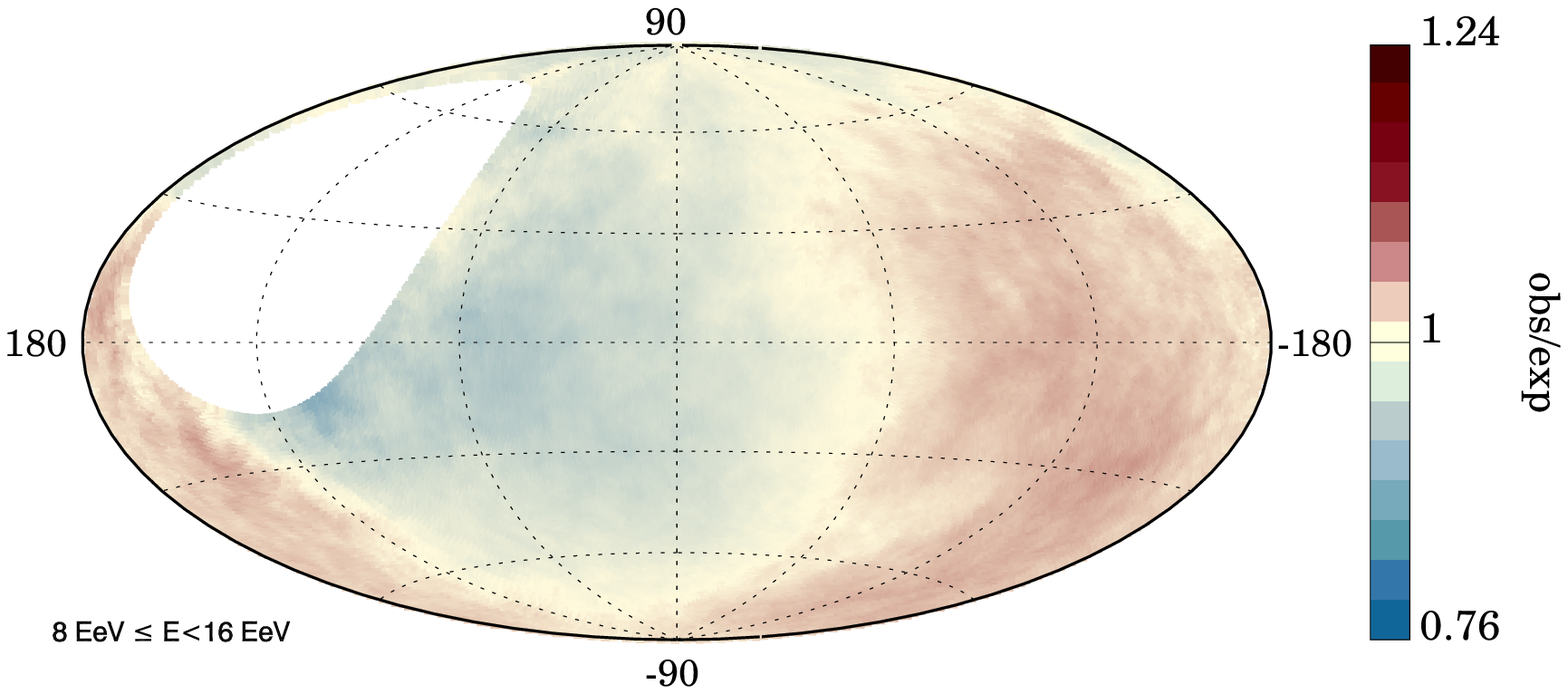}
\includegraphics[scale=.45]{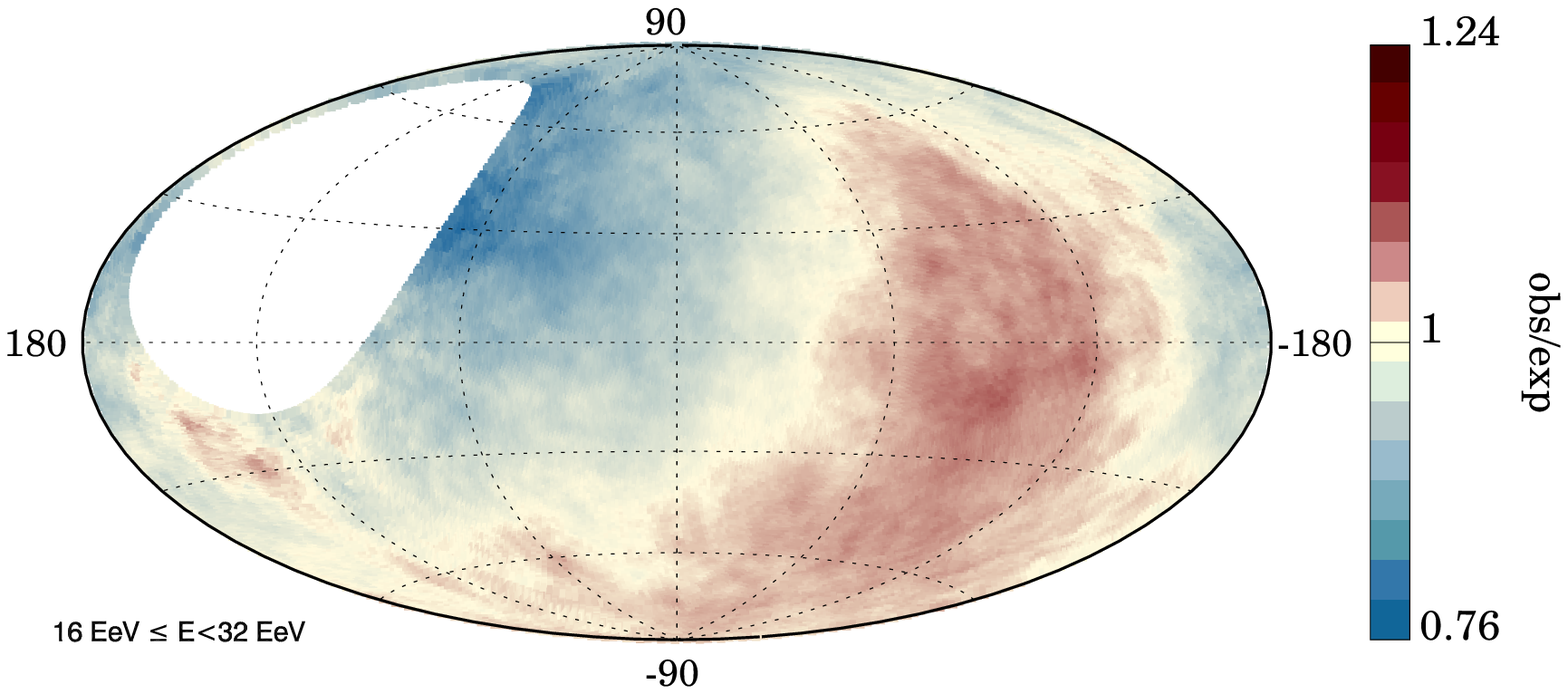}
\includegraphics[scale=.45]{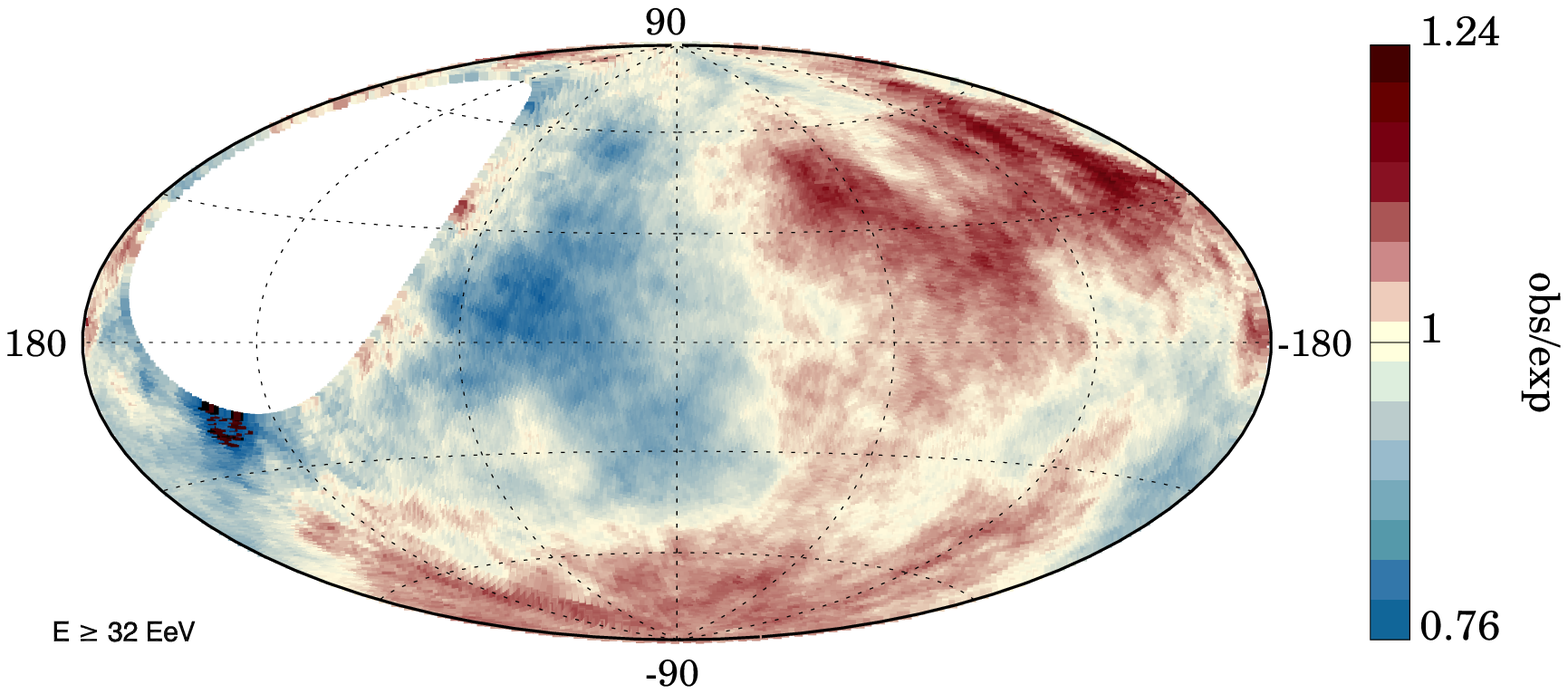}
\caption{Maps in Galactic coordinates of the ratio between the number of observed events in windows of 45$^\circ$ over those expected for an isotropic distribution of arrival directions, for the four energy bins above 4~EeV.}
\label{fig:oea4}
\end{figure}

Figure~\ref{fig:oea4} shows sky maps, in Galactic coordinates, of the ratio between the observed flux and that expected for an isotropic distribution, averaged  in angular windows of 45$^\circ$ radius, for the different energy bins above 4~EeV. The location of the main overdense regions can be observed. Note that the color scale is kept fixed, so as to better appreciate the increase in the amplitude of the flux variations with increasing energies. 

\subsection{Reconstruction of a dipole plus quadrupole pattern}
In order to quantify the amplitude of the quadrupolar moments and their effects on the dipole reconstruction, we assume now that the angular distribution of the CR flux can be well approximated by the combination of a dipole plus a quadrupole. In this case, the flux can be parametrized as
\begin{equation}
\label{phidipquad}
\Phi({\hat u})=\Phi_0\left(1+{\vec d}\cdot {\hat u}+\frac{1}{2}\sum_{i,j}Q_{ij}u_iu_j\right),
\end{equation}
with $Q_{ij}$ being the symmetric and traceless quadrupole tensor.

\begin{table}[ht]
\centering
\caption{Results of the first harmonic in right ascension, separating the events in those arriving from the southern (S) and northern (N) hemispheres.}
\begin{tabular}{c c c c c c c }
\hline\hline
  Energy [EeV] & Hemisphere & $N$ &  $a_1^\alpha$ &  $b_1^\alpha$ & $r_1^\alpha$ & $\varphi_1^\alpha$ [$^\circ$] \\
\hline
 4 - 8  & S & 65,183 & $0.003 \pm 0.005$ & $0.005 \pm 0.005$ & $0.006$ & $60\pm 50$ \\
 & N & 16,518 & $-0.009 \pm 0.011$ & $0.003 \pm 0.011$ & $0.010$ & $160\pm 60$ \\
 \hline
$\geq$ 8  & S & 25,823 & $-0.011 \pm 0.009$ & $0.047 \pm 0.009$ & $0.048$ & $103\pm 10$ \\
 & N & 6,364 & $0.0024 \pm 0.018$ & $0.041 \pm 0.018$ & $0.041$ & $87\pm 25$ \\
\end{tabular}
\label{tab:ns}
\end{table}

The components of the dipole and of the quadrupole can be estimated as in ~\citet{lsa2015}. They are obtained from the first and second harmonics in right ascension and azimuth, given in Tables~\ref{tab:radist} and \ref{tab:phidist}, as well as considering  the first harmonic in right ascension of the events coming from the northern and southern hemispheres separately, which are reported in Table~\ref{tab:ns}. 
 From these results we obtained the  three dipole components and the five independent quadrupole components that are reported in Table~\ref{tab:dipquad}, for the two energy bins [4,\,8]~EeV and $E \ge 8$~EeV.
  The only non-vanishing correlation coefficients between the quantities reported in Table~\ref{tab:dipquad} are $\rho(d_x,Q_{xz})=\rho(d_y,Q_{yz})=0.63$ and $\rho(d_z,Q_{zz})=0.91$. The nine components of the quadrupole tensor can be readily obtained from those  in Table~\ref{tab:dipquad} exploiting the condition that the tensor be symmetric and traceless.
None of the the quadrupole components  is statistically significant and  the reconstructed  dipoles are consistent with those obtained before under the assumption that no higher multipoles are present.
 They are also consistent with results obtained in past analyses in \citet{lsa2015} and \citet{augerta}. Note that allowing for the presence of a quadrupole leads to larger uncertainties in the reconstructed dipole components, specially in the one along the Earth rotation axis due to the incomplete sky coverage present around the North celestial pole. Indeed, in both energy bins the uncertainties in the equatorial dipole components increase by $\sim 30$\% while those on $d_z$ increase  by a factor of about 2.7.

 From the components of the quadrupole tensor it is possible to define an average quadrupole amplitude, $Q\equiv\sqrt{\sum_{ij}Q_{ij}^2/9}$. This amplitude is directly related to the usual angular power-spectrum moments $C_\ell$ through $Q^2=(50/3)C_2/C_0$, and it is hence  a rotationally invariant quantity. From the results given in Table~\ref{tab:dipquad}  one obtains that $Q=0.012\pm 0.009$ for $4\leq E/{\rm EeV}<8$ and $Q=0.032\pm 0.014$ for $E\ge8$~EeV. We note that for isotropic realizations, 95\% of the values of $Q$ would be below 0.037 and 0.060, respectively, showing that the quadrupole amplitude is consistent with isotropic expectations.

\begin{table}[ht]
\centering
\caption{Reconstructed  dipole and quadrupole components in the two energy bins.  The $x$ axis lies in the direction $\alpha=0$.}
\begin{tabular}{c l l }
\hline\hline
 Energy [EeV] & $d_i$ & $Q_{ij}$\\
\hline
4 - 8 & $d_x=-0.005 \pm 0.008$ & $Q_{zz}=-0.01 \pm 0.04$ \\
& $d_y=0.005 \pm 0.008$ & 
$Q_{xx}-Q_{yy}=-0.007 \pm 0.029$ \\
& $d_z=-0.032 \pm 0.024$ & $Q_{xy}=0.004 \pm 0.015$ \\
& & $Q_{xz}=-0.020 \pm 0.019$ \\
& & $Q_{yz}=-0.005 \pm 0.019$ \\
\hline
$\geq 8$ & $d_x=-0.003 \pm 0.013$ & $Q_{zz}=0.02 \pm 0.06$ \\
& $d_y=0.050 \pm 0.013$ & 
$Q_{xx}-Q_{yy}=0.08 \pm 0.05$ \\
& $d_z=-0.02 \pm 0.04$ & $Q_{xy}=0.038 \pm 0.024$ \\
& & $Q_{xz}=0.02 \pm 0.03$ \\
& & $Q_{yz}=-0.03 \pm 0.03$ \\
\end{tabular}
\label{tab:dipquad}
\end{table}

\section{On the dipole uncertainties}\label{systematics}
Let us now discuss the impact of the different systematic effects that we have accounted for. The variations in the array size with time and the atmospheric variations are the two systematic effects that could influence the estimation of the equatorial component of the dipole. Had we neglected the changes in the array size with time it would have changed $d_\perp$, with the dataset considered, by less than $4\times 10^{-4}$, and not performing the atmospheric corrections would have changed $d_\perp$ by less than $10^{-3}$ (the precise amount of the change in these  two cases  depends on the particular phase of $d_\perp$ in each energy bin).
The small values of the effects due to atmospheric corrections and changes in the exposure are mostly due to the fact that for the present dataset they are averaged over a period of more than 12 years.
On the other hand, the tilt of the array and the effects of the geomagnetic field on the shower development can influence the estimation of the North-South dipole component. 
The net effect of including the tilt of the array when performing  observations up to zenith angles  of 80$^\circ$ is to change  $d_z$ by +0.004, which is small since the Observatory site is in a very flat location. The largest effect is that associated to the geomagnetic corrections, which change  $d_z$ by +0.011. Since these corrections  are known with an uncertainty of about 25\% \citep{geomagnetic}, they leave as a remnant a systematic uncertainty on $d_z$ of about 0.003.

A standard check to verify that all the systematic effects that can influence the right-ascension distribution are accurately accounted for,  in particular those arising from atmospheric effects or from the variations in the exposure of the array with time, is to look at the Fourier amplitude at the solar and anti-sidereal frequencies \citep{farley}. No significant physical modulation of cosmic rays should be present at these frequencies for an anisotropy of astrophysical origin. We report in Table~\ref{tab:solas} the results of the first-harmonic analysis at these two frequencies. One can see that the flux modulations at both the solar and anti-sidereal frequencies, having amplitudes with a  sizable chance probability, are in fact compatible with zero  for the two energy ranges considered.

\begin{table}[ht]
\centering
\caption{First-harmonic amplitude, and probability for it to arise as a fluctuation of an isotropic distribution, at the  solar and anti-sidereal frequencies.}
\begin{tabular}{ c  c c  c c  }
\hline\hline
Energy & \multicolumn{2}{c}{solar }  & \multicolumn{2}{c}{anti-sidereal }  \\
$[$EeV$]$ & $r_1$ & $P(\ge r_1)$  & $r_1$ & $P(\ge r_1)$  \\
\hline
4 - 8  & $0.006$ & 0.48 & $0.004$  & 0.76 \\
$\ge 8$  & $0.007$ & 0.69 & $0.011$ &  0.36\\
\end{tabular}
\label{tab:solas}
\end{table}

Regarding the effects of possible systematic distortions in the       zenith-angle distributions, such as those that could arise for instance from a mismatch between the energy calibration of vertical and inclined events, they could affect the dipole components by modifying the quantities $\langle\sin\theta\rangle$ or $\langle\cos\delta\rangle$ entering in Eq.~(\ref{eqdip}).
 Considering for instance the $E\geq 8$~EeV  bin, we note that  for these events $\langle \sin\theta\rangle=0.6525$ while the expected value that is obtained from simulations with a dipolar distribution with amplitude and direction similar to the reconstructed one and the same number of events is $\langle \sin\theta\rangle=0.6558\pm 0.0013$ (while an isotropic distribution would lead to a central value $\langle \sin\theta\rangle=0.6565$). If the difference between the observed  and the expected values of  $\langle \sin\theta\rangle$, which is less than 1\%, were attributed to systematic effects in the zenith distribution, the impact that this would have on the inferred dipole component $d_z$ would be negligible in comparison to its statistical uncertainty, which is about 50\%. Similarly, the value of the average declination cosine in the data is $\langle \cos\delta\rangle=0.7814$, while that expected for the inferred dipole obtained through simulations is $0.7811\pm 0.0013$, showing that possible systematic effects on $d_\perp$ arising from  this quantity are even smaller. This is a verification that the method adopted is largely insensitive to possible systematic distortions in the zenith or declination distribution of the events.

\section{Discussion}

\begin{figure}[h]
\centering
\includegraphics[scale=0.65,angle=0]{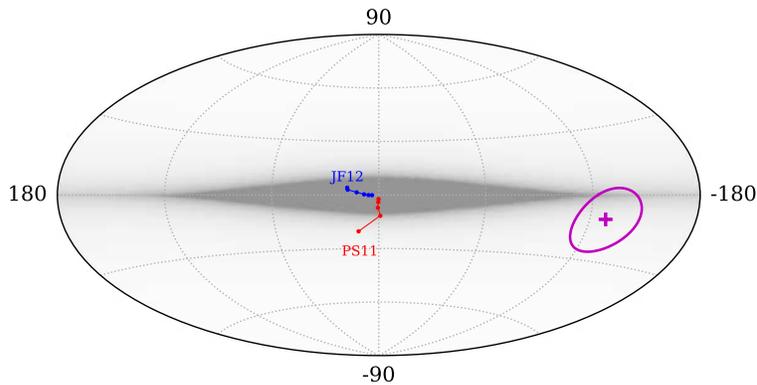}
\caption{Map in Galactic coordinates of the direction of the dipolar component of the flux for different particle rigidities for cosmic rays coming from Galactic sources and propagating in the Galactic magnetic-field model of \citet{JF12} (blue points) and the bisymmetric model of \citet{pshirkov} (red points). The points show the results for the following rigidities: 64~EV, 32~EV, 16~EV, 8~EV, 4~EV and 2~EV (with increasing distance from the Galactic center). We also show in purple the observed direction of the dipole for $E \ge 8 $~EeV and the $68\%$~CL region for it. The background in gray indicates the integrated matter density profile assumed for the Galactic source distribution \citep{weber10}.}
\label{fig:dipgal}
\end{figure}

The most significant anisotropy in the distribution of cosmic rays observed in the studies performed above 4~EeV is the large-scale dipolar modulation  of the flux at energies above 8~EeV. The maximum of this modulation
lies in Galactic coordinates at $(l,b)=(233^\circ,-13^\circ)$, with an  uncertainty of about 15$^\circ$. This is $~125^\circ$ away from the Galactic center direction, indicating an extragalactic origin for these ultrahigh-energy particles. 
As examples of the large-scale anisotropies expected from a Galactic CR component, we show in Fig.~\ref{fig:dipgal} the direction of the dipole that would result for cosmic rays coming from sources distributed as  the luminous matter in the Galaxy, taken as a bulge and an exponential disk modeled as in  \citet{weber10}. The CRs are propagated through the Galactic magnetic field, described with the models proposed in \citet{JF12} and  \citet{pshirkov}, for different values of the CR rigidity, $R=E/eZ$ (with $eZ$ the charge of the CR nucleus). The results are obtained by actually backtracking the trajectories of antiparticles leaving the Earth \citep{thielheim} from a dense grid of equally spaced directions and obtaining the associated weight for each direction by integrating the matter density along their path through the Galaxy \citep{karakula72}. We obtain in this way an estimation of the flux that would arrive at the Earth from a continuous distribution of sources isotropically emitting cosmic rays and with a density proportional to that of the luminous matter. The points in the plot indicate the direction of the reconstructed  dipolar component of the flux maps obtained. The direction of the resulting   dipoles  lie very close to the Galactic center for particles with the highest rigidities considered, and  as the rigidity decreases they slowly move away from it towards increasing Galactic longitudes (closer to the direction of the inner spiral arm which is at $(l,b)\simeq(80^\circ,0^\circ)$). Note that at 10~EeV  the inferred average values  of the CR charges is $Z\sim 1.7$ to 5, depending on the hadronic models adopted for the analysis, while in the lower energy bin the inferred charges are actually smaller \citep{sein}, justifying the range of rigidities considered.
The resulting dipole directions obtained in these Galactic scenarios are quite different from the dipole direction observed above 8~EeV, clearly showing that in a standard scenario the dominant contribution to the dipolar modulation at these energies cannot arise from a Galactic component.  Besides the dipole direction, let us note that the amplitude of the dipole (and also the amplitudes of the quadrupole) turns out to be large in the models of purely Galactic cosmic rays depicted in the figure. In particular, we find that $d>0.8$ for all the rigidities considered, showing that the dominant component at these energies needs to be much more isotropic, and hence of likely extragalactic origin.

Regarding the possible origin of the dipolar CR anisotropy, we note that the relative motion of the observer with respect to the rest frame of cosmic rays is expected to give rise to a dipolar modulation of the flux, known as the Compton--Getting effect \citep{cg}. For particles with a power-law energy spectrum d$\Phi/{\rm d}E\propto E^{-\gamma}$, the resulting dipolar amplitude is $d_{\rm CG}=(v/c)(\gamma+2)$, with $v/c$ the velocity of the observer normalized to the speed of light. In particular, if the rest frame of the cosmic rays were the same as that of the cosmic microwave background, the dipole amplitude would be $d_{\rm CG} \simeq 0.006$ \citep{ks06}, an order of magnitude smaller than the observed dipole above 8~EeV. Thus, the Compton--Getting effect is predicted to give only a sub-dominant contribution to the dipole measured for energies above 8~EeV.

\begin{figure}[t]
\centering
\includegraphics[scale=0.75,angle=-90]{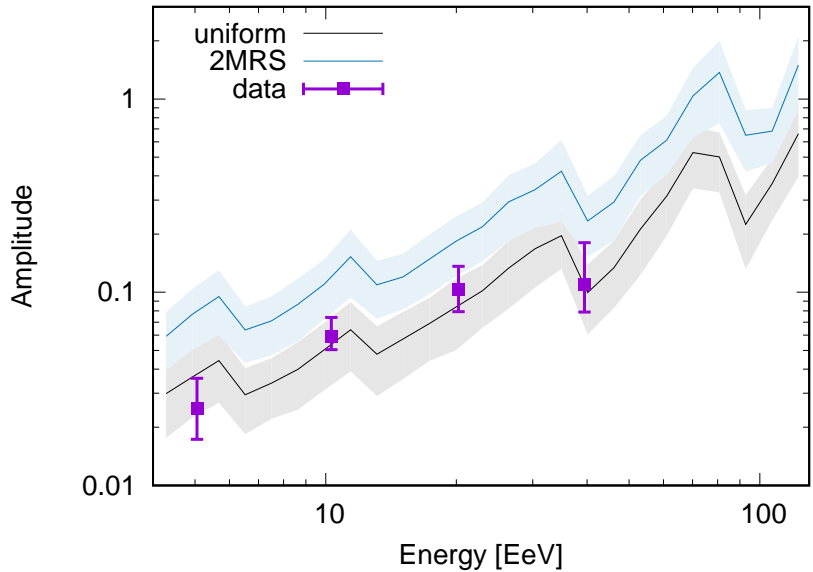}
\caption{Comparison of the dipole amplitude as a function of energy with predictions from models \citep{hmr15} with mixed composition and a source density $\rho = 10^{-4} \,{\rm Mpc}^{-3}$. Cosmic rays are propagated in an isotropic turbulent extragalactic magnetic field with rms amplitude of 1~nG and a Kolmogorov spectrum with coherence length equal to 1~Mpc (with the results having only mild dependence on the magnetic-field strength adopted). The gray line indicates the mean value for simulations with uniformly distributed sources, while the blue one shows the mean value for realizations with sources distributed as the galaxies in the 2MRS catalog.  The bands represent the dispersion for different realizations of the source distribution. The steps observed reflect the rigidity cutoff of the different mass components. }
\label{fig:dvsEdif}
\end{figure}

Plausible explanations for the observed dipolar-like distribution include the diffusive propagation from the closest extragalactic source(s) or that it be due to the inhomogeneous distribution of the sources in our cosmic neighborhood \citep{gi80,be90,hmr14,hmr15}. The expected amplitude of the resulting dipole depends in these cases mostly on the number density of the source distribution, $\rho$, with only a mild dependence on the  amplitude of the  extragalactic magnetic field.\footnote{This is because, as the value of the magnetic field is increased, for any given nearby source closer than the magnetic horizon its contribution to the CR density increases as it gets enhanced by the diffusion while, on the other hand, the value of the dipolar component of its
anisotropy decreases in such a way that both changes compensate for each other to a large extent.} For homogeneous source distributions with $\rho \sim (10^{-5}-10^{-3})$ Mpc$^{-3}$, spanning the range between densities of galaxy clusters, jetted radio-galaxies, Seyfert galaxies and starburst galaxies, the dipole amplitude  turns out to be at the level of few percent at $E\sim 10$ EeV, both for scenarios with light \citep{hmr14}
and with mixed CR compositions \citep{hmr15}. A density of sources smaller by a factor of ten leads on average to a dipolar amplitude larger by approximately a factor of two.  An enhanced anisotropy could result if the sources were to follow the inhomogeneous distribution of the local galaxies, with a dipole amplitude larger by a factor of about two with respect to the case of a uniform distribution of the same source density. 
The expected behavior is exemplified in Figure~\ref{fig:dvsEdif} where we have included the observed dipole amplitude values together with the predictions from
\citet{hmr15} for a scenario with five representative mass components (H, He, C, Si and Fe) having an $E^{-2}$ spectrum with a sharp rigidity cutoff at 6~EV and adopting a source density $\rho = 10^{-4}\, {\rm Mpc}^{-3}$ (ignoring the effects of the Galactic magnetic field).
The data show indications of a growth in the amplitude with increasing energy that is similar to the one obtained  in the models. Note that  this kind of scenario is also in line with the composition favored by Pierre Auger Observatory data \citep{combinedfit}.

\begin{figure}[t]
\centering
\includegraphics[scale=0.75,angle=-0]
{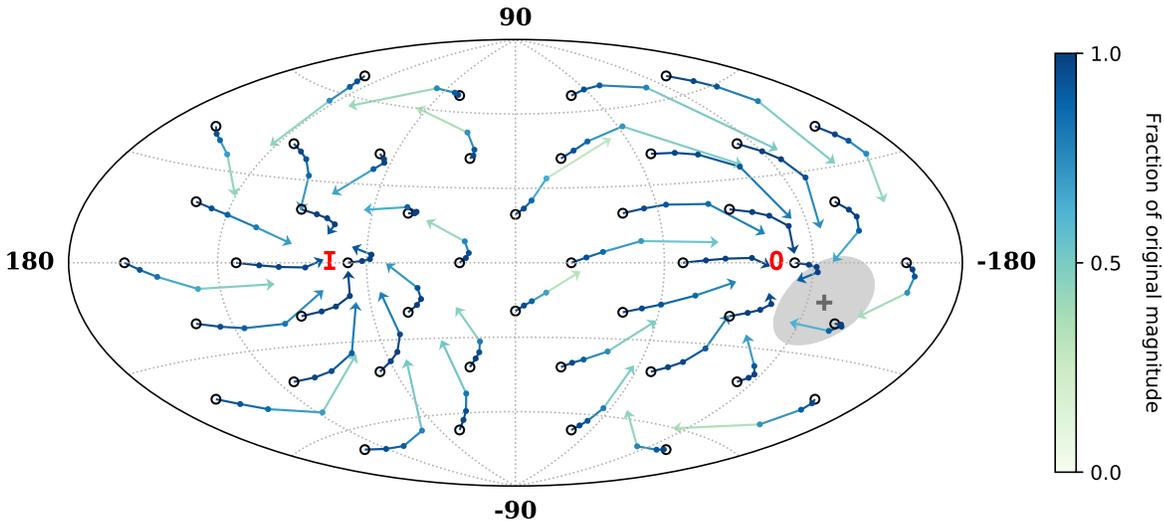}
\caption{Change of the direction of the dipolar component of an extragalactic flux after traversing the Galactic magnetic field, modeled as in \citet{JF12}. We consider a grid (black circles) corresponding to the directions of a purely dipolar flux outside the Galaxy. Points along the lines indicate the reconstructed directions for different values of the particle rigidity: 32~EV, 16~EV, 8~EV and, at the tip of the arrow, 4~EV, respectively. The line color indicates the resulting fractional change of the dipole amplitude. The observed direction of the dipole for energies $E\ge 8$~EeV is indicated by the gray cross, with the shaded area indicating the $68\%$~CL region. The labels {\bf I} and {\bf O} indicate the directions towards the inner and outer spiral arms, respectively. }
\label{fig:dipdirdef}
\end{figure}

Regarding the direction of the dipolar modulation, it is important to take into account the effect of the Galactic magnetic field on the trajectories of extragalactic cosmic rays reaching the Earth.\footnote{These deflections can not only lead to a significant change in the dipole direction and in its amplitude, but they also generate some higher order harmonics even if pure dipolar modulation is only present outside  the Galaxy \citep{hmr10}.} The facts that the Galactic magnetic field model is not well known and that  the CR composition is still uncertain make it difficult to infer the dipole direction associated to the flux outside the Galaxy from the measured one. As an example, we show in Fig.~\ref{fig:dipdirdef} the change in the direction of an originally dipolar distribution after
traversing a particular Galactic magnetic field, modeled in this example following \citet{JF12}. The arrows start in a grid   of initial directions for the dipole outside the Galaxy and indicate   the dipole directions that would be reconstructed  at the Earth  for different CR rigidities. The points along the lines indicate the directions for rigidities of 32~EV, 16~EV, 8~EV and the tip of the arrow those for 4~EV, respectively. We see that after traversing the Galactic magnetic field the extragalactic dipoles originally pointing in one half of the sky, essentially that of positive Galactic longitudes,  tend to have their directions aligned closer to the inner spiral arm, at $(l,b) \simeq (80^\circ,0^\circ)$ (indicated with an {\bf I} in the plot). On the other hand, those originally pointing to the opposite half tend to align their directions towards the outer spiral arm, at $(l,b) \simeq (-100^\circ,0^\circ)$ (indicated with an {\bf O} in the plot). The  measured dipole direction for $E\ge 8$~EeV is indicated with the shaded area and one can see that it lies not far from the outer spiral arm direction.  The line color shows the resulting suppression factor of the dipole amplitude after the effects of the Galactic magnetic field deflections are taken into account.
Qualitatively similar results, showing a tendency for the direction of the dipolar component to align with the spiral arm directions, are also obtained when adopting instead the Galactic magnetic field from \citet{pshirkov}.

 The detection of large-scale anisotropies could open the possibility to jointly probe the distribution of UHECR sources and that of extragalactic magnetic fields \citep{si04}.  In particular, the growth of the dipole with energy is reproduced in the scenarios considered in  \citet{wk17}, \citet{di18} and  \citet{hackstein18}, which further investigate the expected strength of the quadrupolar moments, none of which is found to be significant in our study. In  \citet{wk17} actually the full angular power spectrum $C_l$ up to $l= 32$  is obtained considering the mixed CR composition scenarios with a common maximum rigidity at the sources that best fit the Pierre Auger Observatory results  \citep{witICRC17}. They found that only for $l = 1$, corresponding to the dipole, is the $C_l$ expected to be greater than the 5$\sigma$ CL range of isotropy when a number of events like that recorded by the Pierre Auger Observatory is considered. In  \citet{di18} the dipole and quadrupole amplitudes are examined under several assumptions on the mass composition, for a scenario of sources distributed as in the 2MASS Galaxy Redshift Catalog. The amplitudes of the dipole moment reported in the present work can be well reproduced in their scenario with intermediate mass nuclei. In  \citet{hackstein18} pure proton or pure iron compositions and different magnetogenesis and source distribution scenarios are considered. For the proton case, the first multipole above 8~EeV is generally lower than the measured value (see also \citet{hack16}), while a value closer to the observed one is obtained for the pure iron case. It is also concluded that UHECR large-scale anisotropies do not carry much information on the genesis and distribution of extragalactic magnetic fields. The dependence of the dipolar anisotropies on the root mean square amplitude and coherence length of a turbulent homogeneous intergalactic magnetic field was studied in  \citet{globus17}, for proton, He and CNO source models. They found that the dipole amplitudes for  $E \ge 8$~EeV turn out to be of the order of the one observed for a range of magnetic-field parameters and their model is consistent with an increase  of the dipole amplitude with energy.  In summary,  the dipolar amplitude mostly depends on the large scale distribution of the sources and their density, but it is not very sensitive to the details of the extragalactic magnetic field. Information on the extragalactic magnetic field parameters may eventually be obtained from the determination of anisotropies on smaller angular scales, for which a larger number of events would be needed.

\section{Conclusions}

We have extended the analysis of the large angular scale anisotropies of the cosmic rays detected by the Pierre Auger Observatory for energies above 4~EeV. The harmonic analyses both in right ascension and in azimuth allowed us to reconstruct the three components of the dipole under the assumption that the higher multipoles are sub-dominant. As already described in \citet{science}, for the bin above 8~EeV the first-harmonic modulation in right ascension has a $p$-value  of $2.6\times 10^{-8}$.
 The amplitude of the three-dimensional reconstructed dipole is $d=0.065^{+0.013}_{-0.009}$ for $E\geq 8$~EeV, pointing towards Galactic coordinates $(l,b)=(233^\circ,-13^\circ)$, suggestive of an extragalactic origin for these CRs.  For $4\,{\rm EeV}\leq E<  8$~EeV the dipole amplitude is $d=0.025^{+0.010}_{-0.007}$. Allowing for the presence of a quadrupolar modulation in the distribution of arrival directions, we determined here the three dipolar  and the five quadrupolar components in the [4,\,8]~EeV and $E\ge8$~EeV bins. None of the quadrupolar components turned out to be statistically significant and the dipolar components are consistent with the dipole-only results. 

 We also split the bin above 8~EeV into three to study a possible dependence of the dipole with energy. The direction of the dipole suggests an extragalactic origin for the cosmic-ray anisotropies in each energy bin.   We find 
 that the amplitude increases with energy above 4~EeV, with a constant amplitude being disfavored at the 3.7$\sigma$ level. A  growing  amplitude of the dipole with increasing energies is expected due to the smaller deflections suffered by cosmic rays at higher rigidities. The dipole amplitude is also enhanced for increasing energies due to the increased attenuation suffered by the CR from distant sources, which implies an increase in the relative contribution to the flux arising from the nearby sources, leading to a more anisotropic flux distribution.

 Further clues to understand the origin of the UHECRs are expected to result from the study of the anisotropies at small or intermediate angular scales  for energy thresholds even higher than those considered here. Also the extension of the studies of  anisotropies at large angular scales to lower energies may provide crucial information to understand the transition between the Galactic and extragalactic origins of cosmic rays.

\acknowledgments


\section*{Acknowledgments}

\begin{sloppypar}
The successful installation, commissioning, and operation of the Pierre
Auger Observatory would not have been possible without the strong
commitment and effort from the technical and administrative staff in
Malarg\"ue. We are very grateful to the following agencies and
organizations for financial support:
\end{sloppypar}

\begin{sloppypar}
Argentina -- Comisi\'on Nacional de Energ\'\i{}a At\'omica; Agencia Nacional de
Promoci\'on Cient\'\i{}fica y Tecnol\'ogica (ANPCyT); Consejo Nacional de
Investigaciones Cient\'\i{}ficas y T\'ecnicas (CONICET); Gobierno de la
Provincia de Mendoza; Municipalidad de Malarg\"ue; NDM Holdings and Valle
Las Le\~nas; in gratitude for their continuing cooperation over land
access; Australia -- the Australian Research Council; Brazil -- Conselho
Nacional de Desenvolvimento Cient\'\i{}fico e Tecnol\'ogico (CNPq);
Financiadora de Estudos e Projetos (FINEP); Funda\c{c}\~ao de Amparo \`a
Pesquisa do Estado de Rio de Janeiro (FAPERJ); S\~ao Paulo Research
Foundation (FAPESP) Grants No.~2010/07359-6 and No.~1999/05404-3;
Minist\'erio da Ci\^encia, Tecnologia, Inova\c{c}\~oes e Comunica\c{c}\~oes (MCTIC);
Czech Republic -- Grant No.~MSMT CR LTT18004, LO1305, LM2015038 and
CZ.02.1.01/0.0/0.0/16\_013/0001402; France -- Centre de Calcul
IN2P3/CNRS; Centre National de la Recherche Scientifique (CNRS); Conseil
R\'egional Ile-de-France; D\'epartement Physique Nucl\'eaire et Corpusculaire
(PNC-IN2P3/CNRS); D\'epartement Sciences de l'Univers (SDU-INSU/CNRS);
Institut Lagrange de Paris (ILP) Grant No.~LABEX ANR-10-LABX-63 within
the Investissements d'Avenir Programme Grant No.~ANR-11-IDEX-0004-02;
Germany -- Bundesministerium f\"ur Bildung und Forschung (BMBF); Deutsche
Forschungsgemeinschaft (DFG); Finanzministerium Baden-W\"urttemberg;
Helmholtz Alliance for Astroparticle Physics (HAP);
Helmholtz-Gemeinschaft Deutscher Forschungszentren (HGF); Ministerium
f\"ur Innovation, Wissenschaft und Forschung des Landes
Nordrhein-Westfalen; Ministerium f\"ur Wissenschaft, Forschung und Kunst
des Landes Baden-W\"urttemberg; Italy -- Istituto Nazionale di Fisica
Nucleare (INFN); Istituto Nazionale di Astrofisica (INAF); Ministero
dell'Istruzione, dell'Universit\'a e della Ricerca (MIUR); CETEMPS Center
of Excellence; Ministero degli Affari Esteri (MAE); M\'exico -- Consejo
Nacional de Ciencia y Tecnolog\'\i{}a (CONACYT) No.~167733; Universidad
Nacional Aut\'onoma de M\'exico (UNAM); PAPIIT DGAPA-UNAM; The Netherlands
-- Ministry of Education, Culture and Science; Netherlands Organisation
for Scientific Research (NWO); Dutch national e-infrastructure with the
support of SURF Cooperative; Poland -- National Centre for Research and
Development, Grant No.~ERA-NET-ASPERA/02/11; National Science Centre,
Grants No.~2013/08/M/ST9/00322, No.~2016/23/B/ST9/01635 and No.~HARMONIA
5--2013/10/M/ST9/00062, UMO-2016/22/M/ST9/00198; Portugal -- Portuguese
national funds and FEDER funds within Programa Operacional Factores de
Competitividade through Funda\c{c}\~ao para a Ci\^encia e a Tecnologia
(COMPETE); Romania -- Romanian Ministry of Research and Innovation
CNCS/CCCDI-UESFISCDI, projects
PN-III-P1-1.2-PCCDI-2017-0839/19PCCDI/2018, PN-III-P2-2.1-PED-2016-1922,
PN-III-P2-2.1-PED-2016-1659 and PN18090102 within PNCDI III; Slovenia --
Slovenian Research Agency; Spain -- Comunidad de Madrid; Fondo Europeo
de Desarrollo Regional (FEDER) funds; Ministerio de Econom\'\i{}a y
Competitividad; Xunta de Galicia; European Community 7th Framework
Program Grant No.~FP7-PEOPLE-2012-IEF-328826; USA -- Department of
Energy, Contracts No.~DE-AC02-07CH11359, No.~DE-FR02-04ER41300,
No.~DE-FG02-99ER41107 and No.~DE-SC0011689; National Science Foundation,
Grant No.~0450696; The Grainger Foundation; Marie Curie-IRSES/EPLANET;
European Particle Physics Latin American Network; European Union 7th
Framework Program, Grant No.~PIRSES-2009-GA-246806; and UNESCO.
\end{sloppypar}

\end{document}